\documentclass[prl,superscriptaddress,twocolumn,a4paper]{revtex4} 

\usepackage{color,amsmath}
\usepackage{color,subfigure}
\usepackage{bm}
\usepackage{epsfig}
\usepackage{natbib}
\usepackage{graphicx}
\usepackage{amssymb,amsmath,amsbsy,amsgen,amsfonts}
\usepackage{dcolumn}
\usepackage{amsthm}
\usepackage{latexsym}
\usepackage{array}
\usepackage{amstext}
\allowdisplaybreaks[1]
\usepackage{epstopdf} 
\usepackage[colorlinks=true,citecolor=blue,linkcolor=red]{hyperref}

\usepackage[usenames,dvipsnames]{xcolor}

\newcommand{\bra}[1]{\left\langle{#1}\right\vert}
\newcommand{\ket}[1]{\left\vert{#1}\right\rangle}
\newcommand{\ketbra}[2]{|#1\rangle \langle#2|}
\newcommand{\be}{\begin{equation}}
\newcommand{\ee}{\end{equation}}
\newcommand{\ba}{\begin{array}}
\newcommand{\ea}{\end{array}}
\newcommand{\bqa}{\begin{eqnarray}}
\newcommand{\eqa}{\end{eqnarray}}

\graphicspath{{./figs/}}

\newcommand{\blue}[1]{\textcolor{black}{#1}}

\begin{document}

\title{A Quantum Error Correction-Enhanced  Magnetometer \\Overcoming the Limit Imposed by Relaxation}

\author{David A. Herrera-Mart\'{i}}
\affiliation{Racah Institute of Physics, The Hebrew University, Jerusalem 
91904, Givat Ram, Israel}
\author{Tuvia Gefen}
\affiliation{Racah Institute of Physics, The Hebrew University, Jerusalem 
91904, Givat Ram, Israel}
\author{Dorit Aharonov}
\affiliation{School of Computer Science and Engineering, The Hebrew University, Jerusalem 91904, Givat Ram, Israel.}
\author{Nadav Katz}
\affiliation{Racah Institute of Physics, The Hebrew University, Jerusalem 
91904, Givat Ram, Israel}
\author{Alex Retzker}
\affiliation{Racah Institute of Physics, The Hebrew University, Jerusalem 
91904, Givat Ram, Israel}
\date{\today}

\begin{abstract}
When incorporated in quantum sensing protocols, quantum error correction can be used to correct for high frequency noise, as the correction procedure does not depend on the actual shape of the noise spectrum. As such, it provides a powerful way to complement usual refocusing techniques. Relaxation imposes a fundamental limit on the sensitivity of state of the art quantum sensors which cannot be overcome by dynamical decoupling. The only way to overcome this is to utilise quantum error correcting codes. We present a superconducting magnetometry design that incorporates approximate quantum error correction, in which the signal is generated by a two qubit Hamiltonian term. This two-qubit term is provided by the dynamics of a tunable coupler between two transmon qubits. For fast enough correction, it is possible to lengthen the coherence time of the device beyond the relaxation limit.
\end{abstract}

\maketitle

\emph{Introduction ---} Quantum technologies have attracted great attention over the last decade due to the outstanding enhancements derived from the ability to manipulate physical systems to the limit dictated by quantum mechanics. Common to all of these technologies is the necessity to decouple quantum systems from their environment, while maximizing control. In the context of quantum metrology, a single highly coherent probe can be used to measure very weak magnetic fields via Ramsey interferometry, with a sensitivity that scales as $\delta B \propto 1/\sqrt{T \cdot T_2}$ \cite{Itano,huelga}, where $T$ is the total experiment time and $T_2$ is the probe coherence time. Whereas pure dephasing noise can be accounted for by means of refocusing techniques \cite{hahn,viola, viola1,bylander,cai,Facchi,Fanchini,Rabl,Chen,Gordon}, the relaxation-limited coherence time, \emph{i.e.} $T_2 = 2T_1$, is a fundamental limit to sensing. In this article, we propose an experimental setup based on superconducting devices in which $T_1$ can be prolonged while sensing a weak signal.

\emph{Quantum Error Correcting Codes (QECC) and  Sensing---}Quantum codes were originally devised to lengthen coherence times of quantum registers \cite{shor, steane,laflamme,gottesman}, and it was realised that if the noise rate is below a threshold constant, quantum coherence can be maintained for arbitrarily long times \cite{knill, aharonov,kitaev}. Recently it was observed that introducing QECC will increase the sensitivity in different scenarios \cite{ozeri, dur, arrad, kessler}, which is extremely promising for future theoretical and experimental developments. QECCs can be designed to distinguish the error from the signal by probing a specific $n$-qubit interaction, which limits the use of QECC to sense exotic interactions. However, obtaining $n$-qubit Hamiltonian terms out of single body interactions by means of virtual transitions offers no advantage, as the increase in lifetime resulting from error correction is cancelled by a decrease in strength of the effective signal \cite{arrad}.

The smallest operator that exact QECCs can probe is a three body interaction \cite{laflamme, steane}, since these codes correct all single-qubit quantum errors. In order to overcome this restriction, we resort to approximated QECC \cite{leung}, where relaxation errors can be corrected while probing two-qubit interactions, such as the one offered by a flux-dependent tunable coupler. This code is defined by the codewords $\ket{\bar 0}= \ket{\psi^+}_{1,2}\ket{\psi^+}_{3,4}$ and $\ket{\bar 1}= \ket{\psi^-}_{1,2}\ket{\psi^-}_{3,4}$,
%
where $\ket{\psi^\pm} = \frac{1}{\sqrt{2}}(\ket{00} \pm \ket{11})$. This code is stabilised \cite{gottesman} by $\mathcal{S}_4 = \{S_1=\sigma^X_1\sigma^X_2\sigma^X_3\sigma^X_4,S_2=\sigma^Z_1\sigma^Z_2I_3I_4,S_3=I_1I_2\sigma^Z_3\sigma^Z_4\}$
 and 
 its logical operators are $\bar X = \sigma^Z_1I_2\sigma^Z_3I_4$  and $\bar Z = \sigma^X_1\sigma^X_2I_3I_4$. 

We assume that refocusing techniques and qubit design \cite{koch1,schreier, houck, koch2, masluk, rigetti,herrera,geller,pinto, barends,barends2, barends3,chen} can be used to push the system's coherence to the relaxation limit. In this scenario, the signal is measured by performing a Ramsey-type experiment at the logical level\cite{ozeri, kessler}. Whereas the physical qubits decay at any time, performing correction after short enough lapses of duration $\tau_\textrm{EC}$ will reduce the failure probability at the logical level.

\emph{Tunable Coupler---}The fundamental problem behind QECC-enhanced quantum sensing is the engineering of a many-body Hamiltonian term with strength proportional to the signal to be estimated. We now explain how to obtain a two-body Hamiltonian term using a tunable coupler between two off-resonant transmon qubits \cite{geller,pinto}. The coupler's Hamiltonian is $H= g_s\sigma^Z_1\sigma^Z_3 + g'_s (\sigma^+_1\sigma^-_3+ \sigma^-_1\sigma^+_3)$, where the strength of the coupling energies $g_s$ and $g'_s$ have been calculated to be of order 1 and 10 MHz, respectively \cite{geller}. A flux threading through the tunable coupler can be used to bias it at the optimal response point, \emph{i.e.} $\Phi_\textrm{coupler} = \Phi_\textrm{signal} + \Phi_\textrm{bias}$. For very weak signals compared to the Flux Quantum, {\em i.e.} for $\Phi_\textrm{signal}/\Phi_0 \ll 1$, the response to a threading signal flux can be linearised.

As can be seen from rewriting $\sigma^+_1\sigma^-_3 + \textrm{h.c.} = \frac{1}{2}(\sigma^X_1\sigma^X_3 + \sigma^Y_1\sigma^Y_3) = \frac{\sigma^X_1\sigma^X_3}{2}( I - \sigma^Z_1\sigma^Z_3)$, the effect of the flip-flop term 
induces uncorrectable evolution outside of the codespace. To cancel this effect, qubits at both ends of the tunable coupler must be far detuned so that the energy transfer between them is supressed. Incidentally, the reason why a simple SQUID junction cannot be used in order to generate a logical signal is that in the computational basis, the SQUID Hamiltonian is as a flip-flop term, which nevertheless does not preclude other interesting aplications \cite{peropadre}. For a detuning $\Delta$ between qubits 1 and 3, it is then possible to rewrite the tunable coupler interaction as
\be
H_\textrm{signal} =  g_s \sigma^Z_1\sigma^Z_3 + \mathcal{O}(g'^2_s/\Delta),
\ee
up to a known correction \cite{supmat} at the logical level, due to the flip-flop interaction, which is negligible for large detunings. 

\begin{figure}[!t]
\centering
\includegraphics[scale=.12]{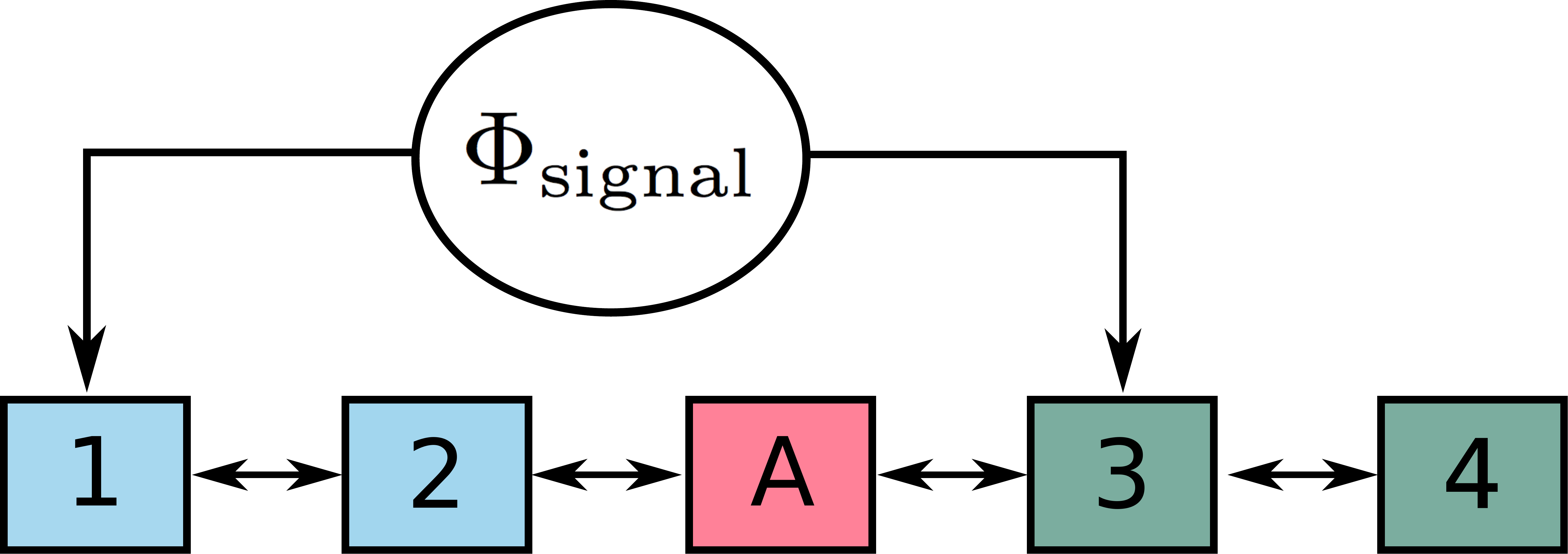}
\caption{Diagrammatic layout of the QECC-enhanced magnetometer. Each superconducting trasmon qubit is represented by a numbered box, and its nearest neighbor tunable coupling (cf. Ref.~\cite{geller, chen, barends2}) is represented by a two-headed arrow. Between qubits 1 and 3, a tunable coupler with a flux-dependent strength $g_s(\Phi_\textrm{signal})$  is interrupted by an inductive loop (cf. \cite{supmat}). This signal is obtained by placing the sensor in proximity to the sample to be measured. Our sensing protocol relies on our ability to prepare the initial state $\ket{\Psi_0} = \ket{\bar 0} = \ket{\psi^+}\ket{\psi^+}$, which is left to precess according to $ H_\textrm{signal}=g_\textrm{s}\sigma^Z_1\sigma^Z_3$. Then, by measuring frequency of the oscillations one is able to infer the value of the magnetic flux threading the coupler.}
\label{physical}
\end{figure}

\emph{Noise and Error Correction ---}The setting presented in Fig.~\ref{physical} bears many similarities with the layout of Ref. \cite{barends2}, where extremely fast quantum gates with high fidelity have been demonstrated. In particular, it was found that the lifetimes are limited by decoherence in the devices and not by noise in gates or in the read-out lines. We therefore model each correction operation as a perfect gate followed by single- and two-qubit depolarizing noise with \blue{per-gate error probability $p_\textrm{gate}$.}

Relaxation at a rate $\gamma$ can be generalised to multiple qubits by $\mathcal{R}(\rho) =\sum_{\mathbf{s}\in\{0,1\}^4} K_\mathbf{s} \rho K^\dag_\mathbf{s}$ with  $K_\mathbf{s} = K_{s_1}K_{s_2}K_{s_3}K_{s_4}$, where  $K_1=\sqrt{\gamma\tau_\textrm{EC}} \sigma^-$ describes a decay event, and $K_0=\ketbra{0}{0}$ $+ \sqrt{1-\gamma\tau_\textrm{EC}}\ketbra{1}{1}$ reflects the fact that if no decay occurred, the probability of finding the qubit in the excited state has decreased. There are only five Kraus operators that act on the state to first order, $K_{0000},K_{0001}, \dots, K_{1000}$.

\begin{figure}[!b]
\centering
\includegraphics[scale = 0.15]{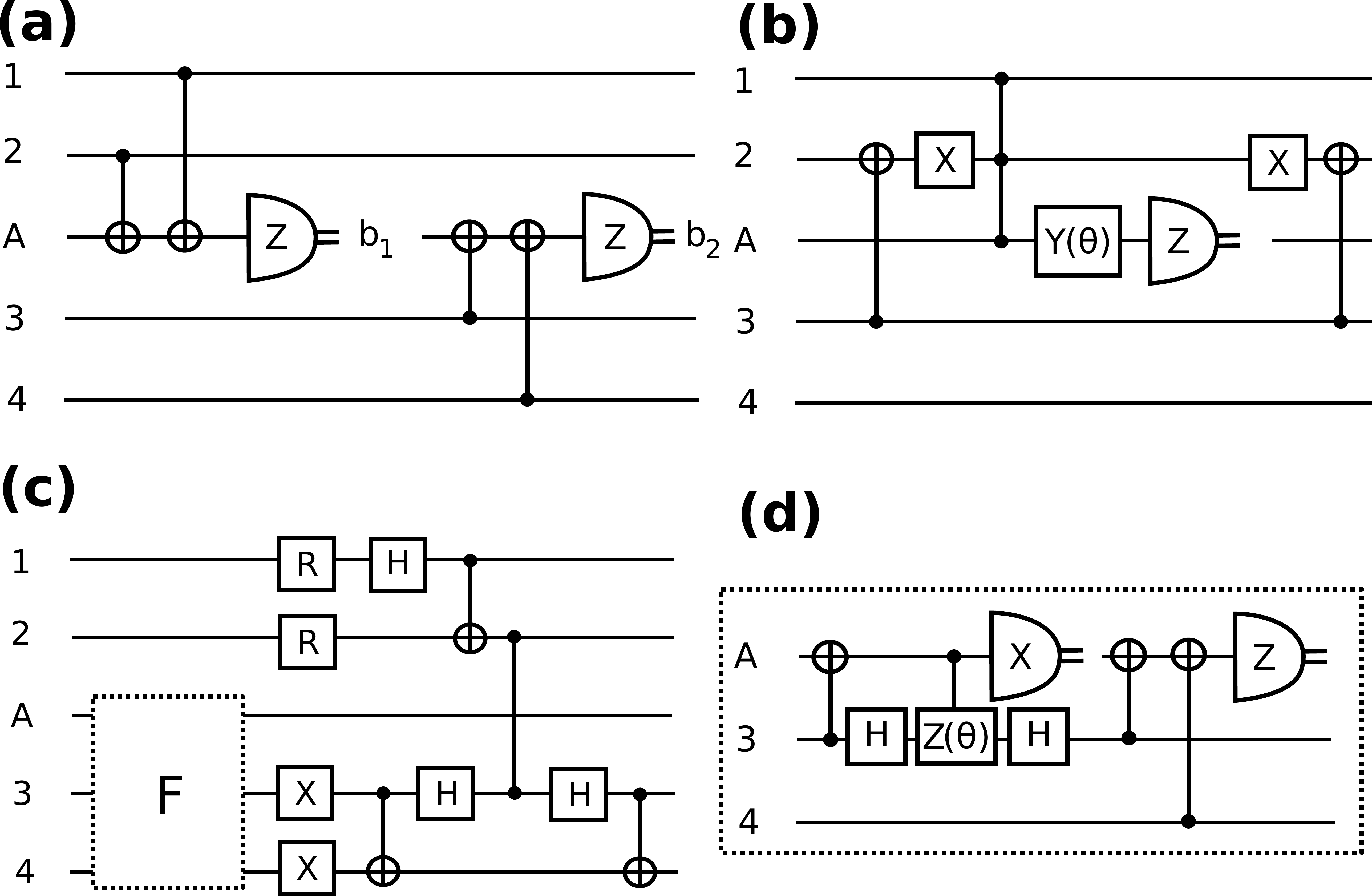}
\caption{Error Correction procedure.{\bf(a)} To detect which of the five first-order operators occurred, a pairwise parity check in the four qubits is stored in two bits $(b_1, b_2)$.{\bf(b)} If $(b_1, b_2)=(0, 0)$, the state is projected in $K^\dag_{0000}\rho K_{0000}$ and this circuit partially undoes the error rotating by  $ \theta=\tan^{-1}((1-\gamma\tau_\textrm{EC})^2)$ in the space of $\{ \ket{0000},\ket{1111} \}$. The ancilla, $A$, is initially prepared in the $\ket{+}$ state and $Y(\theta)$ represents a rotation of $ \theta$ around the $Y$ axis. The Controlled-Controlled-Phase  (CCPhase) gate can be decomposed in a series of five two-qubit gates. This \blue{approximately corrects the} error, up to a known Pauli operator that will depend on the measurement outcome.{\bf(c)}If $(b_1, b_2)=(1, 0)$, then the state is projected in a mixture of $A^\dag_{1000}\rho A_{1000}$ and $A^\dag_{0100}\rho A_{0100}$. A filter is applied in order to regain the relative amplitude between the decayed codewords, followed by a resetting ($R$) of the first two qubits. Qubits 3 and 4 undergo a $X_3X_4$ operation to restore the correct coherence and the following gates are meant to reconstruct the codewords. An equivalent procedure holds for $(b_1, b_2)=(0, 1)$. For $(b_1, b_2)=(1, 1)$, which happens only to second order, no correction is possible and the state is left untouched.{\bf(d)}Filter. The ancilla is initiallised in the $\ket{1}$ state and a conditioned rotation around the Z axis of $ \phi =\cos^{-1}(1-\gamma\tau_\textrm{EC})$ is carried out to transfer part of the amplitude of the $\ket{00}$ state into the odd parity subspace. If even parity is detected, then the correction continues, and it aborts otherwise.}
\label{circuit}
\end{figure}

It was shown in Ref. \cite{leung} that it is possible to outperform standard QECC by relaxing the conditions for error correction. If instead of demanding exact correction for a given error channel, we demand that the approximate QECC retrieves the correct state up to first order in the error probability, then small codes exist that can approximately correct for errors \cite{supmat}. Our error correction protocol $\mathcal{C}$, subsumed in Fig.~\ref{circuit}, ensures that the fidelity of the corrected state is one up to second order, $\mathcal{F} = \textrm{tr}(\mathcal{C}\circ\mathcal{R}(\rho)\rho) \geq 1 - \mathcal{O}((\gamma\tau_\textrm{EC})^2)$, even if the code is not exact. Depending on the outcome of the parity measurement ( Fig.~\ref{circuit}(a)), one of the five possible approximate correction operations is applied ( Fig.~\ref{circuit}(b) and its permutations, and  Fig.~\ref{circuit}(c)). Each of these corrections can be done using single- and two-qubit gates and restores the original state to second order \cite{supmat}. Each SWAP gate needed to perform non-nearest neighbors gates adds an overhead equivalent to three Controlled-Phase gates (of duration $40$ns each \cite{barends2}), and measurements are assumed to take $200$ ns \cite{jeffrey}, rendering the correction doable in about $2$ $\mu$s, with less than thirty two-qubit and a few tens of single-qubit gates on average. This is entirely feasible with current technology \cite{barends3}.

\begin{figure}[!t]
\centering
\includegraphics[scale=.15]{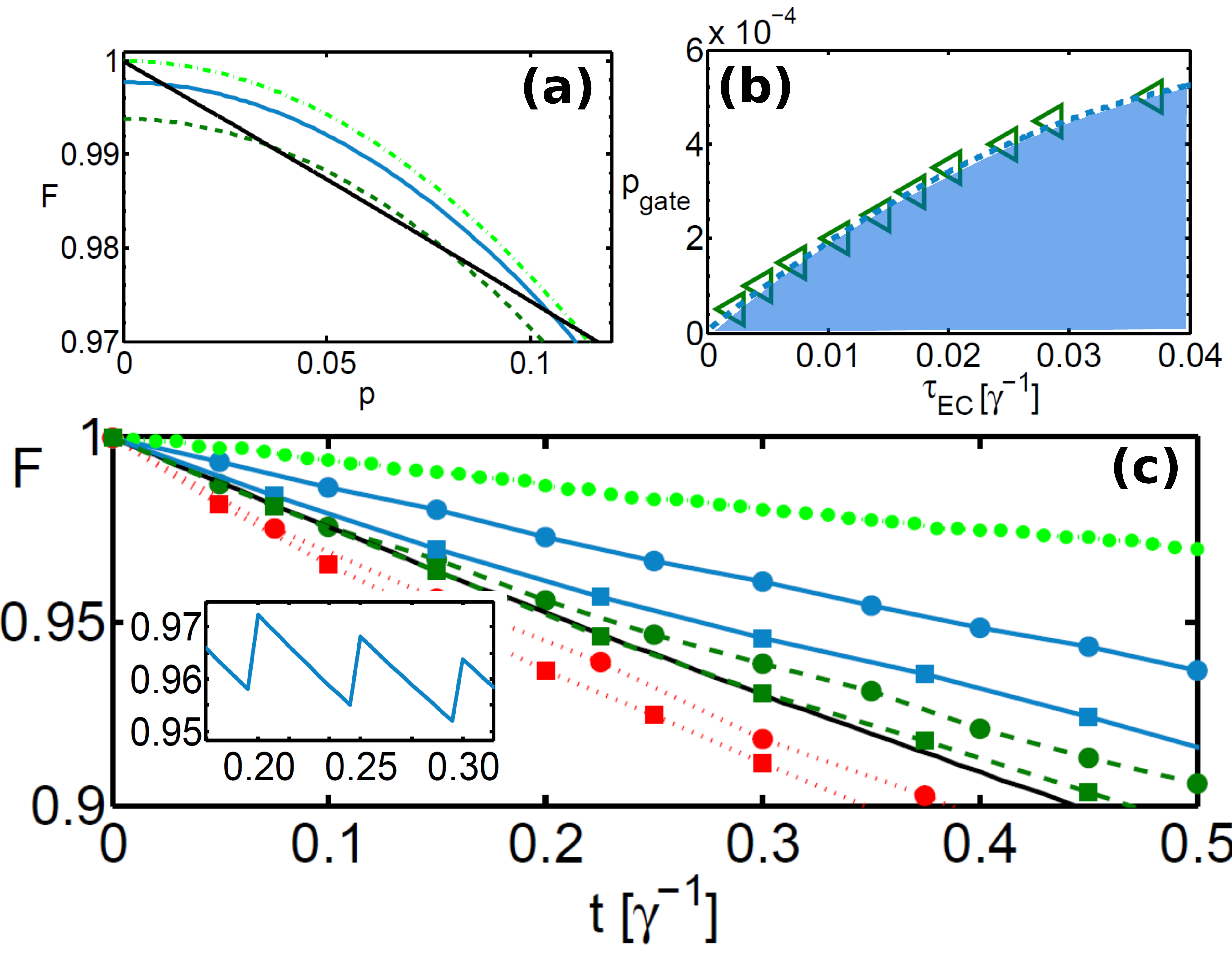}
\caption{{\bf (a)} The fidelity of the initial state $\ket{\bar 0}$ after undergoing relaxation with probability $p$. The solid black line represents the decay of an unencoded probe initially in the superposition state $\ket{+}$. QECC undoes errors to first order in the relaxation probability, as reflected by a quadratic curve (light green, dashed line) in which linear terms do not contribute to fidelity loss. The solid blue line and the dark green dashed line represent the loss of fidelity when the gates in the error correction procedure introduce error with $p_\textrm{gate} = 10^{-2}\%$ and $p_\textrm{gate} = 5\times10^{-2}\%$, respectively.{\bf (b)} Errors introduced in the correction procedure cause a fidelity loss, at a rate depending on the time lapse $\tau_{EC}$. The shaded area below the curve is the area in which there is a sensitivity increase over the unencoded case. {\bf (c)} Decay of the fidelity for a probe initially prepared in the state $\ket{\bar +}$, for three different frequencies of error correction, averaged over $10^4$ runs. Again, the solid black line denotes the evolution of an unencoded probe. The top, light green curve corresponds to $\tau_\textrm{EC}=0.01\gamma^{-1}$ and a gate error of $p_\textrm{gate} = 5\times10^{-3}\%$. The filled circles (squares) denote error correction is carried out every $\tau_\textrm{EC}=0.05\gamma^{-1}$ ($0.075\gamma^{-1}$). The blue solid lines (dashed green) correspond to $p_\textrm{gate} = 10^{-2}\%$($5\times10^{-2}\%$), showing that there is indeed a benefit for high enough gate fidelities and frequencies. The red dotted lines correspond to  $p_\textrm{gate} = 0.1\%$, showing that applying error correction is actually worse than using the unencoded probe.{\bf (inset)} Fine-grained evolution for  $p_\textrm{gate} = 5\times 10^{-2}\%$ and $\tau_\textrm{EC}=0.05\gamma^{-1}$. Between two rounds of error correction, the fidelity decays exponentially.}
\label{threshold}
\end{figure}

\begin{figure}[!t]
\centering
\includegraphics[scale=.20]{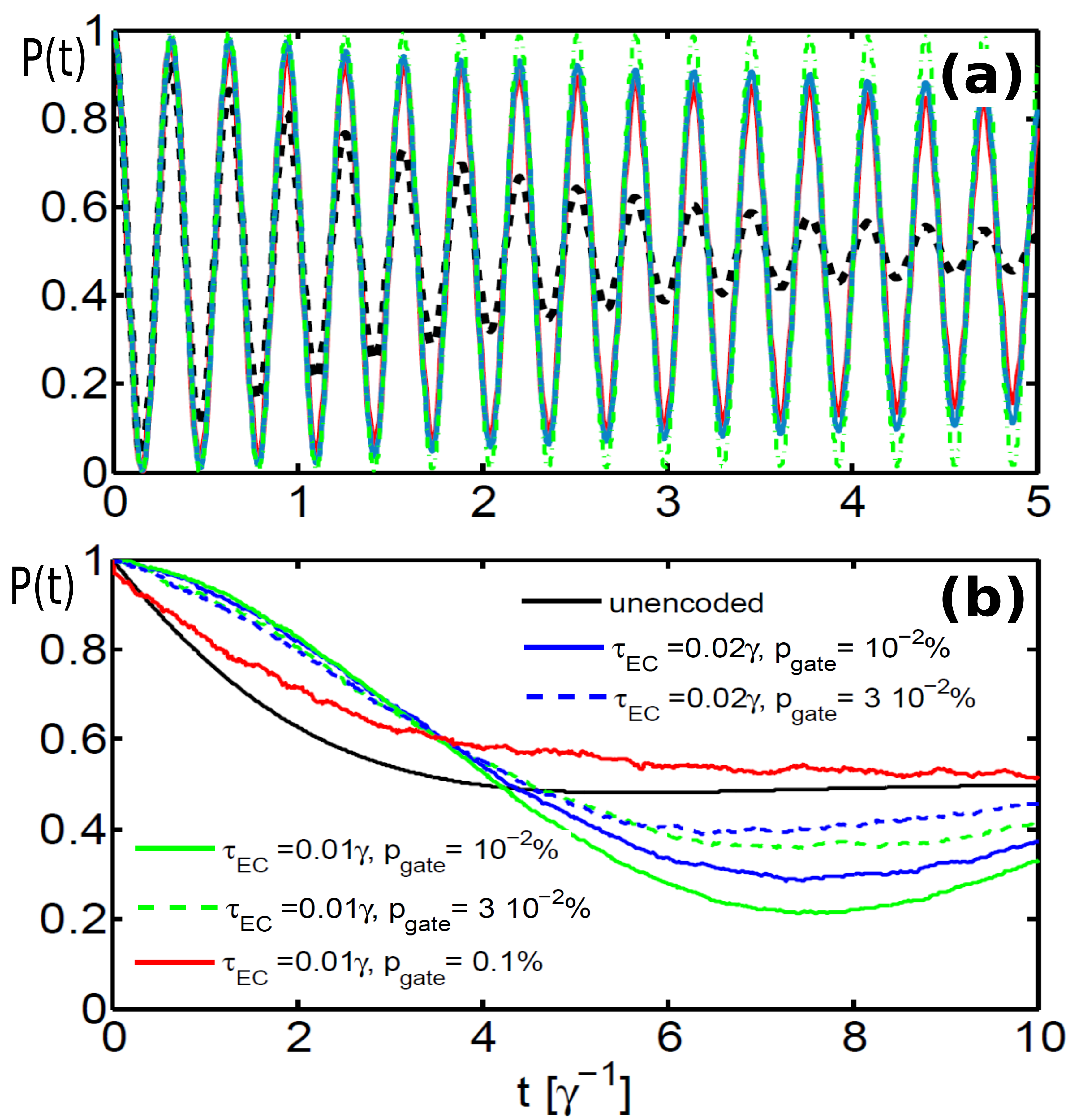}
\caption{Simulation (averaged ever 500 runs) of the population evolution $P(t) = \textrm{tr}(\ketbra{\Psi_0}{\Psi_0}\rho(t))$ in the presence of a signal. {\bf (a)} The signal strength is larger than the decay rate, $g_s = 10\gamma$. The black dashed line shows the evolution of the unencoded probe. The red, blue and dashed green lines represent the evolution with $p_\textrm{gate} = 10^{-2}\%, \tau_\textrm{EC}=10^{-2}\gamma^{-1}$; $p_\textrm{gate}=5\times10^{-3}, \tau_\textrm{EC}=5\times10^{-3}$; and $p_\textrm{gate}=0, \tau_\textrm{EC}=10^{-3}$, respectively. This shows that the fringes contrast can be maintained well beyond the relaxation limit. {\bf (b)} The signal strength is smaller than the decay rate, $g_s = \gamma/5$. Whereas small signals cannot be resolved by a decaying probe, using error correction ensures that the probe remains sensitive to weak signals.} 
\label{flops}
\end{figure}

Below the threshold value depicted in Fig.~\ref{threshold}(a), the error probability at the logical level, $p_\textrm{logical}= 1-\mathcal{F}$,  decreases quadratically as the physical probability  $p = 1-\exp(-\gamma\tau_\textrm{EC}) \approx \gamma\tau_\textrm{EC}$ is reduced since physical errors ocurring at first order are corrected. Importantly, trying to correct for relaxation will introduce errors due to imperfect gates, ancilla preparation and measurements. These errors cannot be accounted for, as a consequence of the four qubit code being too small, and will unavoidably result in a decrease of fidelity. However, larger codes could in principle be used to correct for these errors. In our case, it is still possible to achieve an improvement, provided that the fidelity of the gates is above a threshold that will depend on the time lapse $\tau_\textrm{EC}$. As depicted in Fig.~\ref{threshold}(c), our simulations confirm that lengthening the lifetime beyond the relaxation limit is indeed possible for frequent corrections and sufficiently good gates. In Fig.~\ref{flops} the time evolution of the probe in two scenarios is shown. When the signal is strong compared to the decay rate, it becomes apparent that the contrast can be maintained for times greatly exceeding the relaxation limit. For signals weaker than the inverse lifetime an encoded probe can feel the signal for a time proportional to the inverse effective lifetime, whereas in the unencoded case the signal is rapidly obliterated by the decay.

\emph{Sensitivity Analysis---}The sensitivity of our setup is given by $\delta B = \delta P/|d P/d B|$, \blue{where P is the average value of the measured operator}, and the optimal precision scaling can be analytically calculated to be 
\be
\delta B = \frac{1}{|d g_s/d B|}\sqrt{2e\Gamma^\textrm{eff}/T},
\ee
where $\Gamma^\textrm{eff}$ is the effective noise rate at the logical level, which is estimated analytically \cite{supmat} to be $\Gamma^\textrm{eff} \leq 4\gamma^2\tau_\textrm{EC} + \xi p_\textrm{gate}/\tau_\textrm{EC}$ in the absence of pure dephasing. The parameter $\xi$ is a numerically obtained prefactor which encapsules the collective error of all the gates in the correction protocol \cite{supmat}.

Since $|d g_s/d B| = |d g_s/d\Phi_\textrm{signal}\cdot d\Phi_\textrm{signal}/d B| =|d g_s/d\Phi_\textrm{signal}\cdot A_\textrm{coupler}|$, the larger the area of the coupler, the smaller the magnetic fields that can be measured, at the expense of reducing the spatial resolution of the device. For a linearised response reported in Ref. \cite{geller}, a coupler area size of roughly $100 \mu\textrm{m}^2$, \blue{$T_2 \approx 40 \mu s$}\cite{chen}, and in the absence of error correction, the sensitivity of our design is estimated to be upper-bounded by \blue{$\sim 500 \textrm{pT}\textrm{Hz}^{1/2}$}. It is within reach to improve the circuit parameters to increase the sensitivities by more than one order of magnitude. \blue{This sensitivity compares with those of modern SQUID magnetometers, lying in the $\textrm{nT}\textrm{Hz}^{1/2}$ \cite{vasyukov} and $\textrm{fT}\textrm{Hz}^{1/2}$ \cite{xia} range, depending on application and bandwith. The magnetometer reported in Ref.\cite{hartridge} operates in the Josephson dispersive regime and its sensitivity, not limited by thermal fluctuations, is estimated to fall in the $\textrm{pT}\textrm{Hz}^{1/2}$ range at 600 KHz. We stress that the bandwith in our design is only subject to fluctuations of the biasing flux, since quantum error correction can correct for noise at all frequencies.}

Comparing the energy scales of our system to the typical timescales of interferometry-based sensing schemes with trapped ions, we see that the ratio between the Hamiltonian strengths $ \sim |dg_s/d\Phi_\textrm{signal}| A_\textrm{coupler}/\mu_B$ compensates for the fact that the ions hyperfine levels have longer lifetimes than {\em Xmon} circuits \cite{barends, kotler}, by a factor of $\sim10^5$, and places \blue{our design} as a candidate for the determination of frequency standards. Nitrogen Vacancy centres in diamond are another promising platform for sensing, with sensitivities in the $\textrm{nT}\textrm{Hz}^{1/2}$ range and high spatial resolution \cite{pham,maze,balasubramanian}. Due to the lack of a tunable many-body signal term, it is currently unclear how to supplement this architecture with quantum error correction.

\emph{Considerations on Pure Dephasing---}We have identified two sources of dephasing noise against which our four qubit code is ineffective. The logical information is therefore vulnerable to these errors and must be protected using other methods.

First, one approximation we have taken is that $\tau_\textrm{EC}$ can be arbitrarily short, increasing therewith sensitivity arbitrarily far beyond the relaxation limit. A consequence of $\tau_\textrm{EC}$ being finite, however, is an uncorrectable dephasing caused by uncertainty about when exactly decay happened between two consecutive rounds of correction. In the time lapse between a decay and its correction, the probe evolves outside the logical subspace. As a result, averaging over many realisations of the experiment effectively randomises the accumulated signal. This problem is general in QEC-assisted metrology, yielding a decay rate of $\left( g_s\tau_\textrm{EC}/\hbar \right)^2 \gamma$ \cite{arrad}, which can be mitigated by performing corrections extremely fast.

Second, for sensing AC magnetic fields $B = |B|\cos(\omega_B t)$, applying decoupling pulses at the frequency of the alternating signal will refocus noise due to a fluctuating bias with correlation times longer than $\omega^{-1}_B$. This is especially effective against low-frequency noise, and can indeed be used to push $T_2$ times to the relaxation limit. Importantly, given that these gates can be done very fast, this implies achievable bandwiths of up to several hundred $\textrm{MHz}$.

\emph{Summary \& Outlook---}We suggested a superconducting circuit design to measure magnetic fields with a precision that is not relaxation-limited, by incorporating error correction into the sensing protocol. Correcting at a sufficiently high rate and gate fidelity can increase the lifetime by several orders of magnitude, and it seems possible to probe beyond the femto Tesla regime in the future. For gate speeds and fidelities demonstrated in \cite{barends2, barends3}, we estimate each round of error correction to be achieved in $\sim2 \mu s$. Current {\em Xmon} lifetimes are in excess of $40 \mu s$, meaning that the needed frequencies for error correction are achievable (see Fig.~\ref{threshold}). Therefore, the only remaining impediment are gate fidelities, which should increase by an order of magnitude in order to observe an enhancement. This opens up the possibility to perform quantum metrology in a fault-tolerant manner, that is, probing signals at the logical level while fighting general quantum noise induced by the environment as well as by the correction procedure.

\begin{acknowledgements}
The authors acknowledge discussions with Martin Plenio, Pedram Roushan and Michael Geller. DAHM. acknowledges partial support from a Hebrew University Fellowship, AR. acknowledges support of the Israel Science Foundation (Grant No. 039-8823), the European Commission (Grant No. 323714), DIADEMS Project and the European CIG Grant No. 321798. DA. acknwoledges support from ERC Grant No. 030-8301.
\end{acknowledgements}

\newpage
\onecolumngrid
\appendix

\section{Supplemental Material for \\ A Quantum Error Correction-Enhanced  Magnetometer Overcoming the limit imposed by Relaxation}

\section{Circuit Diagram}

\begin{figure}[!b]
\centering
\includegraphics[scale=.03]{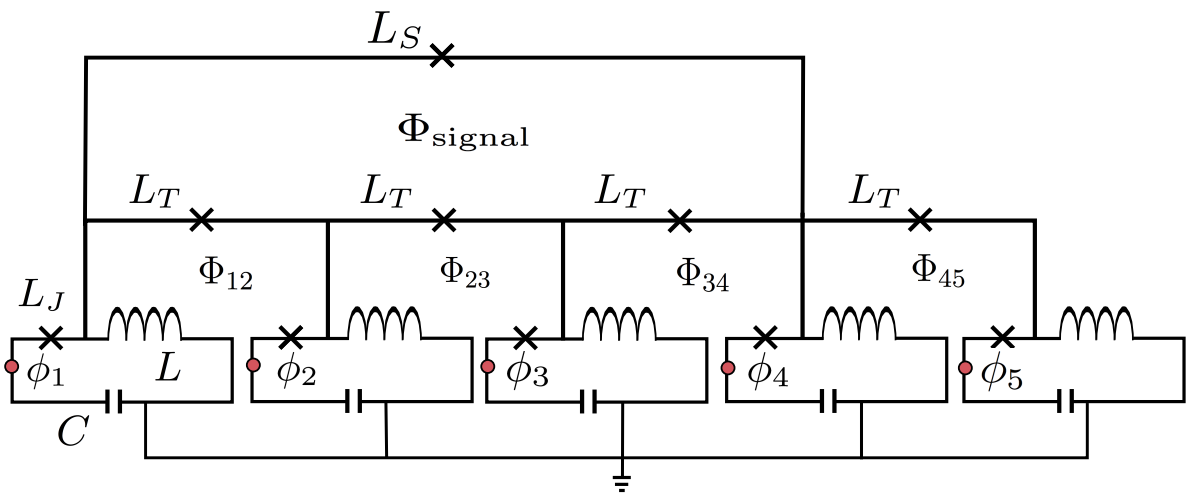}
\caption{Detailed diagram for the protected magnetometer. Each qubit state is physically localized at the red dots. Other active nodes in this representation have no capacitive grounding (they are ``massless") and remain in the instantaneous ground state of the system, and can therefore be eliminated from the dynamics (cf Ref.~\cite{geller}). For each qubit, $L$, $C$ and $L_J$ are taken to be identical (only shown in qubit 1). Although not shown in the circuit, it is assumed that each $L_J$ is flux-tunable. $L_T$ ($L_S$) denotes the Josephson inductance for the nearest-neighbor (signal) Hamiltonian strength. The fluxes $\Phi_{i,i+1}$ are used to enact two-qubit gates needed for error correction. $\Phi_\textrm{signal} = |B|A_\textrm{coupler}$ is the magnetic flux signal.}
\label{layout}
\end{figure}

In Ref. \cite{geller,pinto} a detailed derivation of the Hamiltonian for a tunable coupler is provided. Here we outline how this is done, and further explain how to use the tunable coupler to derive a logical signal Hamiltonian.

Transmon qubits are operated in the phase regime, where the superconducting phase of the aggregate in the island is well-defined. Joining two transmon qubits by means of pure inductively elements opens the door to a flux-dependent current flow between the superconducting islands. Current is the electrical circuit counterpart of mechanical force and, in the case of transmons, this means that an external flux biasing the inductive shunt between the transmon circuits will result in a coupling between the qubits encoded in their eigenbasis. The range of values for maximal and minimal values of the coupling depend on qubit design, as does the values of the external flux for which the coupling vanishes.

In Fig.~\ref{layout} we show in a detailed layout of the magnetometer design. Note that we have chosen a fully numerical notation for the circuits, which is better suited to denote qubit-qubit interaction. Each qubit is inuductively coupled to its nearest neighbors through a tunable coupler, which is needed for syndrome measurement and error correction, and qubits 1 and 4 (qubit 3 in the main text) belong to a further inductive loop closed by $L_S$, which allows for a flux-dependent interaction. This is what causes oscillations in the logical subspace of the code contained in the joint state of circuits 1,2,4 and 5. Because phase is well-defined, it is possible to define the Josephson inductances $L_J$ at each transmon, and the inductances $L_T$ ($L_S$) for the qubit-qubit (signal) interaction.

Let $\phi_i$ and $\phi_j$ denote the phases at both sides of the tunable coupler connecting transmons $i$ and $j$. It is shown in Ref. \cite{geller} how regarding the transmons as anharmonic oscillators allows to write the inductive coupling Hamiltonian as 
$H = g'(\Phi_\textrm{ext})(\sigma^+_i\sigma^-_{i+1}+\sigma^-_i\sigma^+_{i+1})$, where the coupling $g'$ depends only on the external flux and circuit parameters. An additional diagonal $\sigma^Z\sigma^Z$ interaction arises as a result of the level repulsion between the state $\ket{11}$ and the non-computational states $\{ \ket{20},\ket{02}\}$ caused by the transverse coupling, which causes an energy shift of $\ket{11}$ of $\delta E_{11}\approx 2\times(\sqrt{2} g')^2/\alpha$, yielding an effective diagonal interaction $g \approx (g')^2/\alpha$, where $\alpha$ is the anharmonicity of the qubits \cite{geller}. This will be used to generate our logical signal.

In the case of qubits detuned by $\Delta$, the energy differences between states $\{\ket{11},\ket{02}\}$ and $\{\ket{11},\ket{20}\}$ are $\alpha-\Delta$ and $\alpha+\Delta$, respectively. In this case, state repulsion caused by transverse coupling gives the energy shift $\delta E_{11}\approx (\sqrt{2} g')^2/(\alpha+\Delta) + (\sqrt{2} g')^2/(\alpha-\Delta)$, provided that $|\alpha \pm \Delta|\gg g'$. The case of $\Delta \approx \pm\alpha$ would cause strong hybridization of states $\ket{11}$ and $\{\ket{20},\ket{02}\}$, and should be avoided. This fact will be used later to obtain a signal Hamiltonian.

In terms of Pauli operators, the total Hamiltonian for a tunable coupler can be written as  $H=g'(\Phi_\textrm{ext})  (\sigma^+_i\sigma^-_j+\sigma^-_i\sigma^+_j) +  g(\Phi_\textrm{ext}) \sigma^Z_i\sigma^Z_j$. We refer to Ref. \cite{geller} and references therein for a more in-depth analysis of the tunable coupler.

\subsection*{Derivation of the Logical Signal Hamiltonian}

It suffices to assume the coupling mechanism described above to derive a signal Hamiltonian acting on the logical level of the QECC. Starting from the total Hamiltonian for the circuit's device:

\bqa
H = \sum^5_{i=1} \epsilon_i \sigma^Z_i &+& \sum^4_{i=1} g'_{i, i+1}  (\sigma^+_i\sigma^-_{i+1}+\sigma^-_i\sigma^+_{i+1}) +  g_{i, i+1} \sigma^Z_i\sigma^Z_{i+1} \nonumber \\
&&+ g'_s ( \sigma^+_1\sigma^-_4+\sigma^-_1\sigma^+_4) +  g_s \sigma^Z_1\sigma^Z_4
\eqa
where the indices run from qubit 1 to qubit 5 taking into account the ancilla qubit, corresponding to the four qubits in the code plus the ancilla qubit. In the main text, qubit 3 (4,5) corresponds to qubit A (3,4), although here we rename the qubits for ease of notation (see Fig.~\ref{layout}). The couplings between the transmons are tunable and can be used to achieve arbitrary two-qubit interactions, by bringing the qubits in and out of resonance in a controlled manner and changing the current bias in the inductinve loops \cite{chen, geller}. Going to the rotating frame of the qubits we can rewrite the Hamiltonian

\bqa
\tilde H(t) &=& \sum^4_{i=1}g_{i, i+1} \sigma^Z_i\sigma^Z_{i+1} + g'_{i, i+1} ( e^{i2(\epsilon_i-\epsilon_{i+1})t}\sigma^+_i\sigma^-_{i+1}+ e^{-i2(\epsilon_i-\epsilon_{i+1})t}\sigma^-_i\sigma^+_{i+1}) \nonumber\\
&& +  g_s \sigma^Z_1\sigma^Z_{4} + g'_s (e^{i2(\epsilon_1-\epsilon_{4})t}\sigma^+_1\sigma^-_{4}+ e^{-i2(\epsilon_1-\epsilon_{4})t}\sigma^-_1\sigma^+_{4})
\eqa
where the couplings $g_{i,i+1},g'_{i,i+1}$ depend on controlled biases. Our signal will be propotional to the coupings $g_s$ and $g_s'$, which depend on the external flux threading the tunable coupler. This tunability will be used to create the Bell pairs in the code, and to decouple the qubits in the code from the ancilla.

It is clear from the code definition in the main text that we need our signal to be proportional to $\sigma^Z_1\sigma^Z_4$. This can be achieved by ensuring that qubits 1 and 4 are always far-off resonant, which is needed to disregard the flip-flop term in the rotating frame. In this case, after applying the Rotating Wave Approximation, we have:

\bqa
\tilde H(t) &\approx_{RWA}& \sum^4_{i=1}g_{i, i+1} \sigma^Z_i\sigma^Z_{i+1} + g'_{i, i+1} ( e^{i2(\epsilon_i-\epsilon_{i+1})t}\sigma^+_i\sigma^-_{i+1}+ e^{-i2(\epsilon_i-\epsilon_{i+1})t}\sigma^-_i\sigma^+_{i+1}) \nonumber \\
&& +  g_s \sigma^Z_1\sigma^Z_{4} 
\eqa

In this frame, the codewords are:

\bqa
\ket{\tilde{\bar 0}} &=& \frac{1}{2} \left( \ket{\tilde{0000}}+\ket{\tilde{0011}}+\ket{\tilde{1100}}+\ket{\tilde{1111}}\right)\\
\ket{\tilde{\bar 1}} &=& \frac{1}{2} \left( \ket{\tilde{0000}}-\ket{\tilde{0011}}-\ket{\tilde{1100}}+\ket{\tilde{1111}}\right)
\eqa
where the phases coming from the energy splitting are absorbed into the new states $\ket{\tilde{0000}}=\ket{0000}$, $\ket{\tilde{0011}}=e^{-i2(\epsilon_4+\epsilon_5)t}\ket{0011}$, $\ket{\tilde{1100}}=e^{-i2(\epsilon_1+\epsilon_2)t}\ket{1100}$ and $\ket{\tilde{1111}}=e^{-i2(\epsilon_1+\epsilon_2+\epsilon_4+\epsilon_5)t}\ket{1111}$. Again, notice that qubits 4 and 5 correspond to the third and fourth qubit of the code, denoted 3 and 4 in the main text. These phase differences between the states have to be monitored in order to perform error correction in the lab frame.

As an important observation, we should mention that the tunable coupler presented in Ref. \cite{geller} was not optimized to maximize diagonal over transverse coupling. Although we strived to show that our magnetometer suggested prototype is within reach of demonstrated technology, we are convinced that better-suited tunable couplers will not be difficult to design.

\subsection*{Error due to off-resonant transverse coupling}

By detuning qubits $1$ and $4$(1 and 3 in the main text) by $\Delta$, the contribution of the flip-flop interaction is decreased substantially. We now show that it will take the system out of the code space with a probability of $\left( \frac{g'_s}{\Delta} \right)^2$ in leading order. As long as this error is kept lower than the error induced by the gate it will give  a small correction to the final result. The result of this detuning would be an extra term to the Hamiltonian $ \frac{\left(g'_s \right)^2}{\Delta} \left( \sigma_z^1  - \sigma_z^4  \right)$, which in the general case is just a known correction to the phase generated by the signal.

In order to see why we can neglect the action of the transverse coupling in the logical Hamiltonian, consider the simplified version in which the local effect of the flip-flop term is modeled as a $\sigma^X$ operator with strength $g'_s$ acting on qubit 4. Going to the rotating frame of qubit 1 allows us to apply the theory for off-resonant driving, which can be modeled with the Hamiltonian:

\be
H = \frac{\Delta}{2}\sigma^Z_4 + g'_s\sigma^X_4
\ee
where $\Delta = \epsilon_4 - \epsilon_1$ is the detuning between qubits 1 and 4 (from now on we drop the subindex). In this frame, the rotating term cannot be neglected, and acts on qubit 4 by changeing its energy. The dressed states in this frame are:

\bqa
\ket{\lambda_+} &= \frac{1}{\sqrt{1+\eta^2}}(\ket{0} +\eta\ket{1}), \\
\ket{\lambda_-} &= \frac{1}{\sqrt{1+\eta^2}}(-\eta\ket{0} + \ket{1}),
\eqa
with $\eta = g'_s/[\sqrt{(\Delta/2)^2+g^2}+\Delta/2]$. We are interested in the case where the detuning is much larger than the transverse coupling, so that $\eta\approx g'_s/\Delta $ and the energies are $\lambda_\mp = \pm \frac{\Delta}{2}[ 1+ (g'_s/\Delta)^2 + \mathcal{O}((g'_s/\Delta)^4)]$. From here it can be seen that there is a known but negligible correction given by $\sim \frac{g'^2_s}{\Delta}$ to the phase accumulated by the diagonal coupling. However, in the dressed-state basis the relaxation operators will induce some amount of pure dephasing. As can be seen from the expansion of $\sigma_-$:

\bqa
\sigma_- &=& \frac{1}{1+\eta^2}\left(\ketbra{\lambda_+}{\lambda_-} + \eta(\ketbra{\lambda_+}{\lambda_+}-\ketbra{\lambda_-}{\lambda_-})-\eta^2(\ketbra{\lambda_-}{\lambda_+})\right) \nonumber\\
&=& \frac{1}{1+\eta^2}\left(\hat\sigma_- + \eta\hat\sigma^Z-\eta^2\hat\sigma_+\right)
\eqa
where the hats denotes the expansion in the new eigenbasis. Although in the off-resonant case the flip-flop term  (here represented as a energy non-conserving $\sigma^X$ operator for a single qubit, for simplicity) can be treated perturbatively, the new master equation will still have a small, but finite, time-dependent contribution from pure dephasing and excitation processes. In the rotating frame, the master equation for qubit 4 is:

\be
\hat{\dot\rho} \approx \gamma_\downarrow \mathcal{L}[\hat\sigma_-] \hat\rho+ \eta^2\gamma_0 \mathcal{L}[\hat\sigma^Z] \hat\rho + \eta^4\gamma_\uparrow \mathcal{L}[\hat\sigma_+] \hat\rho
\ee
where $\mathcal{L}$ denotes the Lindbladian superoperator and $\{\gamma_\downarrow,\gamma_0,\gamma_\uparrow\}$ correspond to phonomenological relaxation, dephasing and excitation rates. The second term, diagonal in the new eigenbasis, is linked to pure dephasing processes, which are not correctable for the four qubit code. Therefore, the dephasing probability in the eigenbasis will be proportional to $\eta^2\approx(g'_s/\Delta)^2$ in leading order, which should be smaller than or at least comparable to our threshold gate error, taken here to be $p_E \approx 5\times10^{-2}\%$ (see Figs. 3-4 in the main text). For a coupling of $g'_s\approx1\textrm{MHz}$, which we take from \cite{geller}, a detuning of no more than $500 \textrm{MHz}$ would suffice. Therefore we can see that neglecting flip-flop terms is a good approximation, and we expect it to remain so even for more substantial transverse coupling, provided that the qubit detuning can be increased.

Importantly, a substantial diagonal coupling is retained for detuned qubits, given that $|\alpha \pm \Delta|\gg g'$. For example, for a detuning twice the qubit anharmonicity $\Delta = 2\alpha$, the coupling strength would be $\sim66\%$ of the resonant case, whereas the transverse signal would be greatly suppressed.

\section{Approximate Quantum Error Correction}

A sufficient condition for correctability is given by the expression $\bra{c_i}E^\dag_\mu E_\nu\ket{c_j} = \delta_{ij}\delta_{\mu\nu} p_\mu$, where $\ket{c_k}$ are codewords of a given code and $E_\alpha$ are the correctable errors. This condition ensures that each error deforms each codeword by the same amount and moreover keeps them orthorgonal. In the codespace $\Pi_\mathcal{C} = \sum \ketbra{c_i}{c_i}$, these errors have a polar decomposition $E_\alpha\Pi_\mathcal{C} = \sqrt{p_\alpha}U_\alpha\Pi_\mathcal{C}$ where $U_\alpha$ is a unitary transformation, giving $\bra{c_i}U^\dag_\mu U_\nu\ket{c_j} = \delta_{ij}\delta_{\mu\nu}$. Approximate quantum error correction amounts to relaxing the correctability condition to $\bra{c'_i}U^\dag_\mu U_\nu\ket{c'_j} = (\delta_{ij} + \mathcal{O}(\epsilon^{t+1}))\delta_{\mu\nu}$, with $\ket{c'_j}$ belonging to an approximate code. This means that the corrected codewords are not recovered exactly, but only to some given order $t+1$ in the error probability $\epsilon$, which is fine since the aim of quantum codes correcting for $t$ errors is to reduce the effect to the $t+1$th order.  Clearly, if one demands perfect correctability then the approximate conditions reduce to the exact conditions. 

Relaxation processes can be modeled in the operator sum representation, using the amplitude damping channel. There are five operators that can be accounted for using the QECC introduced in the main text, $K_{0000},K_{1000},K_{0100},K_{0010}$ and $K_{0001}$.  A central observation \cite{leung} is that all of these operators, when acting upon the codespace $\Pi_\mathcal{C}$, have a polar decomposition of the form $K_\mathbf{s}\Pi_\mathcal{C} = U_\mathbf{s}(\sqrt{\lambda_\mathbf{s}} I + Q_\mathbf{s})\Pi_\mathcal{C}$, where $\lambda_\mathbf{s}$ is the smallest eigenvalue of $\Pi_\mathcal{C} K^\dag_\mathbf{s}K_\mathbf{s}\Pi_\mathcal{C}$ and $Q_\mathbf{s} = \sqrt{\Pi_\mathcal{C} K^\dag_\mathbf{s}K_\mathbf{s}\Pi_\mathcal{C}} - \sqrt{\lambda_\mathbf{s}}\Pi_\mathcal{C}$ is a semipositive operator that quantifies the unrecoverable distortion inflicted in the codewords. In other words, each of these operators can be decomposed in a non-trivial rotation on the codespace, \textit{i.e.} $\bra{\bar{0}}U_\mathbf{s}\ket{\bar{1}}\neq 0$, preceded by a dilation. Only the part  proportional to $\sqrt{\lambda_\mathbf{s}} I$ can be undone, whereas the one proportional to $Q_\mathbf{s}$ is not recoverable. 

The fidelity of the corrected state is given by the expression:

\bqa
F &\geq& \textrm{min}_{\ket{c'_j}\in \mathcal{C}} \sum_{\mu\in M_\textrm{correctable}} |\bra{c'_j}U^\dag_\mu K_\mu\ket{c'_j}|^2\\
&=&\textrm{min}_{\ket{c'_j}\in \mathcal{C}} \sum_{\mu\in M_\textrm{correctable}} |\bra{c'_j}(\sqrt{\lambda_\mu} I + U^\dag_\mu Q_\mu)\ket{c'_j}|^2\\
&=& \sum_{\mu\in M_\textrm{correctable}}  \lambda_\mu,
\eqa
where $M_\textrm{correctable}=\{K_{0000},K_{1000},K_{0100},K_{0010}, K_{0001}\}$ denotes the set of correctable errors in the operator sum representation. The circuits in Fig. 2 are aimed at performing the operations $U^\dag_\alpha$ using only single- and two-qubit operations.

As an explicit example, if the state $\ket{\bar 0}$ undergoes the action of $K_{0000}$,  the resulting (unnormalised) state is $\ket{0000} + (1-\gamma \tau_\textrm{EC}) \left ( \ket{1100} + \ket{0011} \right)+ (1-\gamma \tau_\textrm{EC})^2\ket{1111}$, which can be brought to $(1-\gamma \tau_\textrm{EC}) \ket{\bar 0}$ up to second order, via unitary operations. In order to apply this correction the precise knowledge of the decay rate is needed. It can be shown  (\cite{leung}) that the uncorrectable part of the error only contributes to second order in $\gamma\tau_\textrm{EC}$ to the fidelity loss. Error correction thus amounts to detecting which of the five possible first-order processes occurred, and then undoing the unitary part of the error.

\section{Effective Decoherence Rate and Sensitivity}

In order to develop further insight into the decoherence process at the logical level, we derived a formula for the effective decoherence rate analytically by diagonalizing the action of the relaxation Liouivillian in the operation sum representation, parametrized by the deacay probability $\gamma\tau_\textrm{EC}$, followed by an error correction procedure intruducing noise at an effective probability $p^\textrm{eff}_\textrm{gate} = \xi p_\textrm{gate}$, and observed that the loss coherence in the logical subspace is given by the rate:

\be
\Gamma^\textrm{eff} \leq 4\gamma^2\tau_\textrm{EC} + \xi p_\textrm{gate}/\tau_\textrm{EC},
\ee
where $\xi \approx  8.4(\pm3)$ has been estimated from numerical simulations. 
As can be gleaned from the plot below, for low enough gate error, it is possible to reduce the decoherence rate by increasing the error correction frequency. Whereas for very high fidelity gates, the effective decoherence rate can be reduced several orders of magnitude, above a certain value of the gate error probability, no improvement is possible.

\begin{figure}[!h]
\centering
\includegraphics[scale=.35]{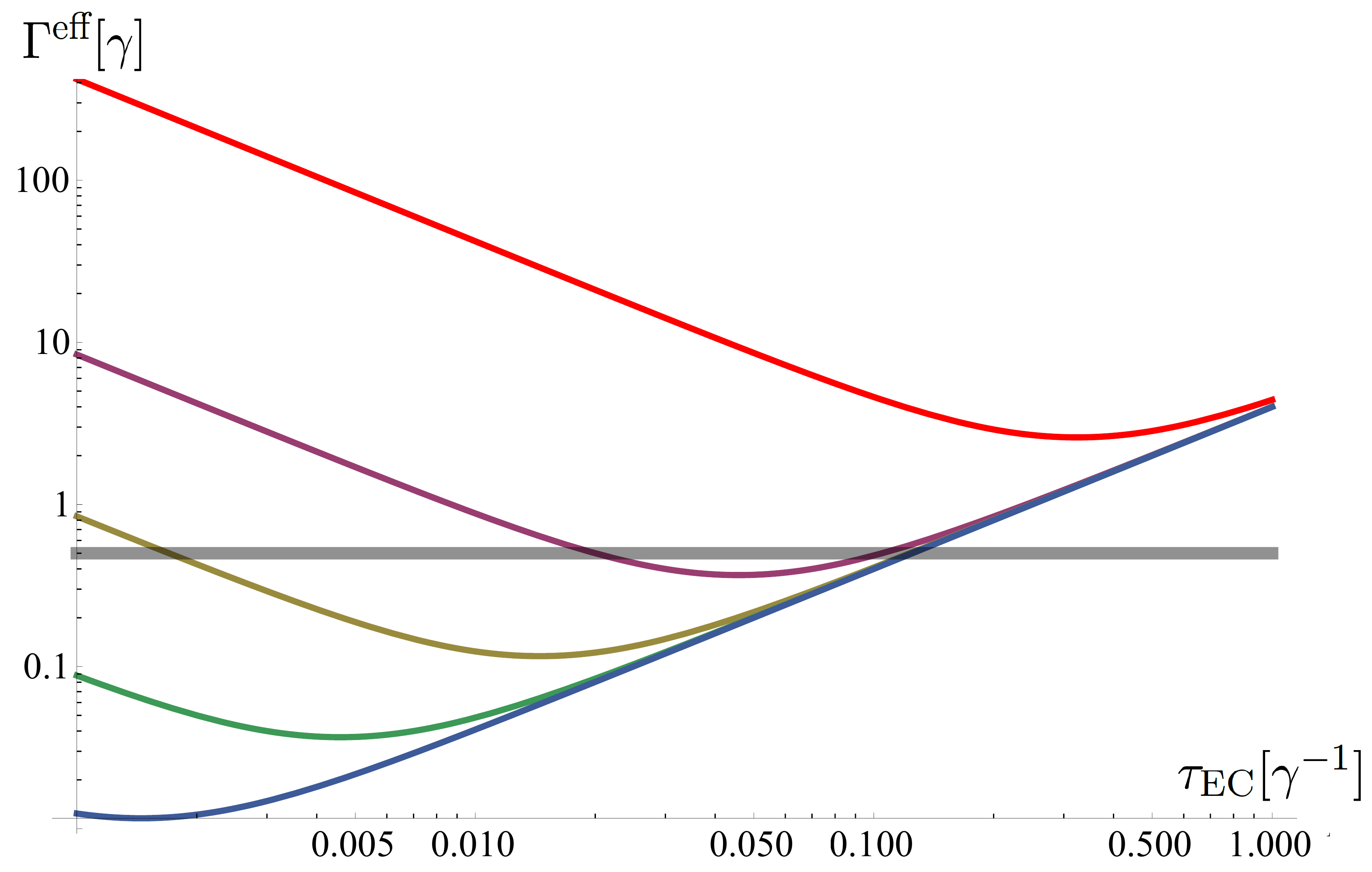}
\caption{Log-log plot of the effective decoherence rate $\Gamma^\textrm{eff}$ versus the error correction time $\tau_\textrm{EC}$ for different values of the gate error probability $p_\textrm{gate}$. The grey line denotes the bare unencoded case, with decay rate $\gamma/2$. In decreasing value, these errors are $p_\textrm{gate}=5\times 10^{-2}, 10^{-3}, 10^{-4}, 10^{-5}$ and $10^{-6}$. }
\end{figure}

These results are in good quantitative agreement with the numerical estimates presented in the main text. However, note that even for a relaxation-limited probe, there exist other decoherence channels at the logical level, such as pure dephasing due to finite $\tau_\textrm{EC}$ and dephasing due to flutuations in the biasing flux, which can be dealt with dynamical decoupling pulses.

The resolution of the logical Ramsey experiment is obtained by the relation $\delta B = \delta P / |d P/d B|$, where $P(t)$ is the probability of measuring an optimal logical operator after an evolution of time $t$. The fluctuations of P come given by $\delta P = \sqrt{(1-P)P/N}$, where N = T/t is the number of experimental points and t (T) is the evolution (total experiment) time. It can be shown (\cite{huelga}) that, in the presence of coherence damping, the optimal evolution time is given by $t^* = 2e/\gamma$, where $\gamma$ is the rate at which the off-diagonal elements vanish. In this case, the optimal resolution is given by 

\be
\delta B = \frac{h}{|dg_s/d\Phi_\textrm{signal}|A_\textrm{coupler}}\sqrt{2e\Gamma^\textrm{eff}/T},
\ee
where we have taken the derivative of $P(t) = \frac{1+ e^{-\Gamma^\textrm{eff}t}\cos(2g_s t)}{2}$ with respect to $|B|$ and used the optimal time $t^*$.

In the case where $\Gamma^\textrm{eff}\rightarrow 0 $, then the sensitivity will scale proportionally to $1/T$, which is the main goal of incorporating QECC within sensing protocols.

\end{document}